\documentclass[twocolumn,showpacs,pra]{revtex4}
\usepackage{amssymb}
\usepackage{makeidx}
\usepackage{amsmath}
\usepackage{graphicx}
\usepackage{dcolumn}
\usepackage{bm}

\begin{document}

\title{Multiphoton interaction of a qutrit with single-mode quantized field
in the ultrastrong and deep strong coupling regimes}
\author{H.K. Avetissian$^{1}$}
\author{A.K. Avetissian$^{1}$}
\author{G.F. Mkrtchian$^{1}$}
\author{O.V. Kibis$^{2}$}
\affiliation{$^{1}$ Centre of Strong Fields Physics, Yerevan State University, 1 A.
Manukian, Yerevan 0025, Armenia}
\affiliation{$^{2}$ Department of Applied and Theoretical Physics, Novosibirsk State
Technical University, Karl Marx Avenue 20, 630073 Novosibirsk, Russia}

\begin{abstract}
We consider multiphoton dynamics of a quantum system composed of a
three-state atom (a qutrit) and a single-mode photonic field in the
ultrastrong and deep strong coupling regimes, when the coupling strength is
comparable to or larger than the oscillator energy scale. We assume a qutrit
to be in a polar-$\Lambda $ configuration in which two lower levels have
mean dipole moments. Direct multiphoton resonant transitions revealing
generalized Rabi oscillations, collapse, and revivals in atomic excitation
probabilities for the ultrastrong couplings are studied. In the deep strong
coupling regime particular emphasis is placed on the ground state of
considering system which exhibits strictly nonclassical properties.
\end{abstract}

\maketitle

\ % It is always \today, today,


\section{Introduction}

Quantum dynamic interactions involving a quantum system with a few energy
levels and one or more near-resonant modes of the quantized photonic field
have been extensively studied by means of various models. Such combined
systems are shown to exhibit interesting nonclassical effects, such as the
collapse and revival of the Rabi oscillations of the atomic inversion,
antibunched light, squeezing, and etc. \cite{Shore,RW}. Among these models
the so called Jaynes-Cummings (JC) model \cite{Jaynes}, which describes a
two-level system coupled to a quantum harmonic oscillator (e.g., a single
radiation mode) has many applications in various branches of contemporary
physics ranging from quantum optics/informatics \cite{Mand,Scully,INF} to
condensed matter physics \cite{Dots,CirQED,Kib}.

With the three-state quantum system, depending on the levels' linkages, one
can enrich the conventional JC model including new effects connected with
the quantum interference effects. With respect to simple two-level systems,
in the three-level system an extra level can be used for effective
manipulation of remaining two levels or so called qubits. Otherwise,
three-state quantum system as a whole can be used as a unit for storing
quantum information. In the latter case as a unit of quantum information
stands for qutrit, which has several specific features providing significant
improvements over qubits for several quantum protocols \cite{LL}. The
various cases of three-state atoms coupled to a quantized field have been
treated by many authors (see \cite{Shore} and references therein). The
considered linkages are the ladder ($\Xi $), the vee ($V$), and the lambda ($%
\Lambda $). As has been shown in Refs. \cite{M1,M2}, there is another
three-state configuration -one can refer it as a $\Gamma $ configuration,
where multiphoton transitions in the quantum dynamics of the system
subjected to a classical radiation field are very effective compared to the $%
\Xi $, $V$, and $\Lambda $ configurations. In this case lower level is
coupled to an upper level which in turn is coupled to an adjacent level. If
the energies of excited states in $\Gamma $ configuration are enough close
to each other then by the unitary transformation the problem can be reduced
to the polar-$V$ configuration, i.e. with permanent dipole moments in the
excited stationary states. In this context as a known example one can
mention the hydrogen atom in spheric and parabolic \cite{Landau}
coordinates. The inverse with respect to the $\Gamma $ configuration is the $%
L$ configuration, which is unitary equivalent to the polar-$\Lambda $
configuration (see below). Thus, it is of interest to study the interaction
of $L$-type (or $\Gamma $-type) atom with single-mode quantized radiation
field, where new multiphoton effects are expected. Thanks to recent
achievements in Cavity/Circuit Quantum Electrodynamics (QED) \cite{scs} one
can achieve interaction-dominated regimes in which multiphoton effects are
expected. The key parameter to characterize Cavity/Circuit Quantum
Electrodynamics (QED) setups is the vacuum Rabi frequency, which is the
strength of the coupling between the light and matter. Depending on the
magnitude of the vacuum Rabi frequency Cavity/Circuit QED can be divided
into four coupling regimes: weak, strong, ultrastrong, and deep strong. For
the weak coupling, the atom-photon interaction rate is smaller than the
atomic and cavity field decay rates. In this case one can manipulate by the
spontaneous emission rate compared with its vacuum level by tuning discrete
cavity modes \cite{3}. In the strong coupling regime, when the
emitter--photon interaction becomes larger than the combined decay rate,
instead of the irreversible spontaneous emission process coherent periodic
energy exchange between the emitter and the photon field in the form of Rabi
oscillations takes place \cite{4}. In the ultrastrong coupling regime, the
emitter--photon coupling strength is comparable to appreciable fractions of
the oscillator frequency \cite{scs,5}. In this regime new nonlinear \cite{NL}
and multiphoton \cite{M3} phenomena are visible that are not present in the
weak or strong coupling regimes. If the emitter--photon coupling strength is
increased even further it becomes larger than the oscillator frequency. This
regime, usually referred to as deep strong coupling regime, opens up new
possibilities for matter--photon manipulations in the quantum level \cite{6}%
. The main candidate for achieving deep strong coupling regime is the
circuit QED setups \cite{CirQED} where one can realize artificial atoms with
desired configuration \cite{7,77}. The main advantage of these atoms over
natural ones is the additional control associated with the tunability of
almost all parameters. In particular, one can realize three-state $\Delta $%
-atom \cite{7}, opening possibilities for many quantum optics phenomena with
superconducting circuits.

In the present paper we consider a quantum system composed of a three-level
artificial atom (qutrit) and a single-mode photonic field, i.e. harmonic
oscillator, in the ultrastrong and deep strong coupling regimes.\textrm{\ }%
We assume a qutrit to be in a polar-$\Lambda $ configuration in which two
lower levels have mean dipole moments. We consider direct multiphoton
resonant transitions for the ultrastrong coupling regime. In the deep strong
coupling regime particular emphasis is placed on the ground state of the
system. The latter exhibits strictly nonclassical properties, which ensures
controllable implementation of qutrit-oscillator entangled states.

The paper is organized as follows. In Sec. II the model Hamiltonian is
introduced and diagonalized in the scope of a resonant approximation. In
Sec. III we consider temporal quantum dynamics of considered system and
present corresponding numerical simulations. In Sec. IV we present results
of numerical calculations that demonstrate the properties of the system in
the deep strong coupling regime. In particular, we consider quantum features
of the ground state. Finally, conclusions are given in Sec. V.

\section{Basic Hamiltonian and Resonant Approximation}

Let us consider a three-state quantum system or so called qutrit interacting
with the single-mode radiation field of frequency $\omega $. Schematic
illustration of the system under consideration is shown in Fig. 1. We assume
a qutrit to be in a polar-$\Lambda $ configuration in which two lower levels 
$\left\vert g_{1}\right\rangle $ and $\left\vert g_{2}\right\rangle $ with
mean dipole moments are coupled to a single upper level $\left\vert
e\right\rangle $. Other possible three-level scheme is shown in the lower
part of Fig. 1 and one can refer it as a $L$ configuration. In this case
upper level is coupled to an lower level which in turn is coupled to an
adjacent level. For the $L$ configuration the mean dipole moment is zero for
a stationary states. The polar-$\Lambda $\ configuration is unitary
equivalent to a $L$\ configuration. A more in-depth discussion of this point
can be found in Appendix~\ref{sec:L}. The mentioned qutrit configuration can
be realized for example in the symmetric double well potential. The latter
is a frequently appearing structure in the solid-state semiconductor or
superconductor systems \cite{Leg}. In particular, the effective potential
landscape is reduced to a double-well potential for superconducting quantum
interference device loop \cite{SQUID} and three-Josephson junction loop \cite%
{three}.

Here we assume coupling to a bosonic field with the transition selection
rules equivalent the ones for the electric-dipole transitions in usual
atoms. For the artificial atom based on the superconducting quantum circuit
the eigenstates involve macroscopic number of electrons. However, as was
shown in Refs. \cite{7,77} the optical selection rules of the
microwave-assisted transitions in a flux qubit superconducting quantum
circuit are the same as the ones for the electric-dipole transitions in
usual atoms when effective potential landscape is reduced to a symmetric
double-well potential.

Thus, assuming the basis 
\begin{equation}
|g_{1}\rangle =\left( 
\begin{array}{c}
1 \\ 
0 \\ 
0%
\end{array}%
\right) ,|g_{2}\rangle =\left( 
\begin{array}{c}
0 \\ 
1 \\ 
0%
\end{array}%
\right) ,|e\rangle =\left( 
\begin{array}{c}
0 \\ 
0 \\ 
1%
\end{array}%
\right) ,  \label{basis}
\end{equation}%
\begin{figure}[tbp]
\includegraphics[width=.45\textwidth]{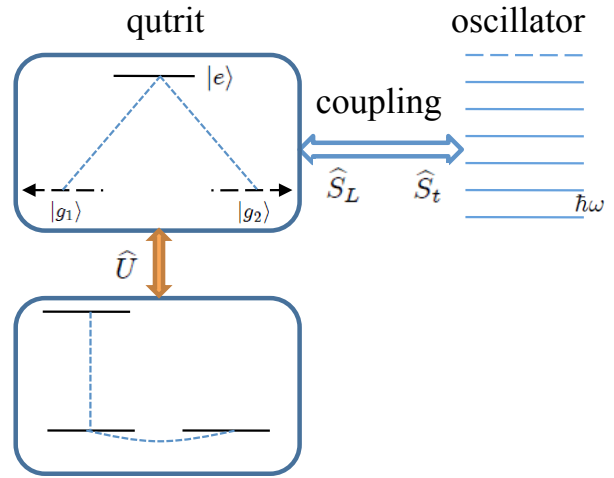}
\caption{ Schematic illustration of the system under consideration. A qutrit
in polar-$\Lambda $ configuration is coupled to a quantized single-mode
field, represented as a harmonic oscillator with characteristic frequency $%
\protect\omega $. Here two lower levels $\left\vert g_{1}\right\rangle $ and 
$\left\vert g_{2}\right\rangle $ with mean dipole moments are coupled to a
single upper level $\left\vert e\right\rangle $. The considered
configuration is unitary equivalent to a $\mathrm{L}$ configuration shown in
the lower part of diagram. In this case upper level is coupled to a lower
level which in turn is coupled to an adjacent level.}
\end{figure}
the Hamiltonian for the polar-$\Lambda $ system coupled to a bosonic field
will be presented in the form: 
\begin{equation*}
\widehat{H}=\hbar \omega \left( \widehat{a}^{+}\widehat{a}+\frac{1}{2}%
\right) +\widehat{H}_{\mathrm{\Lambda }}
\end{equation*}%
\begin{equation}
+\hbar \left( \mu \widehat{S}_{L}+\lambda \widehat{S}_{t}\right) \left( 
\widehat{a}^{+}+\widehat{a}\right) .  \label{H_m}
\end{equation}%
The first term in Eq. (\ref{H_m}) corresponds to the free harmonic
oscillator of frequency $\omega $ (single-mode radiation field). The second
term: 
\begin{equation}
\widehat{H}_{\mathrm{\Lambda }}=\left( 
\begin{array}{ccc}
\varepsilon _{g} & \Delta  & 0 \\ 
\Delta  & \varepsilon _{g} & 0 \\ 
0 & 0 & \varepsilon _{e}%
\end{array}%
\right)   \label{h0}
\end{equation}%
corresponds to the three-level system. Here nondiagonal elements ($\Delta $)
describe transitions between lower laying states (tunnel transition). The
final term in Eq. (\ref{H_m}) gives the interaction between the single-mode
radiation field and qutrit. Creation and annihilation operators, $\widehat{a}%
^{+}$and $\widehat{a}$, satisfy the bosonic commutation rules. The operator 
\begin{equation}
\widehat{S}_{L}=-|g_{1}\rangle \ \langle g_{1}|+|g_{2}\rangle \ \langle
g_{2}|=\left( 
\begin{array}{ccc}
-1 & 0 & 0 \\ 
0 & 1 & 0 \\ 
0 & 0 & 0%
\end{array}%
\right)   \label{SL}
\end{equation}%
is the result of the mean dipole moments in the states of indefinite parity.
The operator 
\begin{equation}
\widehat{S}_{t}=|g_{1}\rangle \langle e|-|g_{2}\rangle \langle e|+\mathrm{%
h.c.}=\left( 
\begin{array}{ccc}
0 & 0 & 1 \\ 
0 & 0 & -1 \\ 
1 & -1 & 0%
\end{array}%
\right)   \label{St}
\end{equation}%
describes transition between excited and lower laying states. At $\mu =0$ we
have the usual Hamiltonian for $\Lambda $ model. At $\lambda =0$ the excited
state is decoupled and after unitary transformation (\ref{A4}) one will
obtain usual Hamiltonian for JC model (including also counter-rotating
terms) with coupling $\hbar \mu $ and atomic energy $2\Delta $. Thus, to
emphasize three-state structure in this paper we will consider the case $%
\left\vert \Delta \right\vert <<\hbar \omega <\varepsilon _{e}-\varepsilon
_{g}$.

At first, we will diagonalize the Hamiltonian (\ref{H_m}) for the moderately
strong couplings, which is straightforward in the resonant case. For the
case $\Delta =0$, the Hamiltonian (\ref{H_m}) can be rewritten in the form: 
\begin{equation}
\widehat{H}=\widehat{H}_{0}+\widehat{V},  \label{dec}
\end{equation}%
where 
\begin{equation}
\widehat{H}_{0}=\widehat{H}_{os}\otimes \widehat{P}_{e}+\widehat{H}%
_{-}\otimes \widehat{P}_{g_{1}}+\widehat{H}_{+}\otimes \widehat{P}_{g_{2}}
\label{H00}
\end{equation}%
represents three non-coupled oscillators. Here $\widehat{P}%
_{g_{1}}=|g_{1}\rangle \ \langle g_{1}|$, $\widehat{P}_{g_{2}}=|g_{2}\rangle
\ \langle g_{2}|$, and $\widehat{P}_{e}=|e\rangle \ \langle e|$ are
projection operators. The excited state is associated to normal oscillator
with the Hamiltonian 
\begin{equation}
\widehat{H}_{os}=\hbar \omega \left( \widehat{a}^{+}\widehat{a}+\frac{1}{2}%
\right) +\varepsilon _{e},  \label{Hph}
\end{equation}%
while two others are associated to position-displaced oscillators 
\begin{equation}
\widehat{H}_{-}=\hbar \omega \left( \widehat{a}^{+}\widehat{a}+\frac{1}{2}%
\right) +\varepsilon _{g}-\hbar \mu \left( \widehat{a}^{+}+\widehat{a}%
\right) ,  \label{Ha}
\end{equation}%
\begin{equation}
\widehat{H}_{+}=\hbar \omega \left( \widehat{a}^{+}\widehat{a}+\frac{1}{2}%
\right) +\varepsilon _{g}+\hbar \mu \left( \widehat{a}^{+}+\widehat{a}%
\right) .  \label{Hb}
\end{equation}%
The interaction part 
\begin{equation}
\widehat{V}=\hbar \lambda \widehat{S}_{t}\left( \widehat{a}^{+}+\widehat{a}%
\right)  \label{h2}
\end{equation}%
in Eq. (\ref{dec}) couples $\widehat{H}_{os}$ with $\widehat{H}_{-}$ and $%
\widehat{H}_{+}$. Hamiltonians (\ref{Hph}), (\ref{Ha}), and (\ref{Hb}) admit
exact diagonalization. It is easy to see that the corresponding eigenstates
are%
\begin{eqnarray}
|e,N^{(os)}\rangle &\equiv &|e\rangle \otimes |N\rangle ,  \notag \\
|g_{1},N^{(-)}\rangle &\equiv &|g_{1}\rangle \otimes e^{(\mu /\omega )(\hat{a%
}^{\dag }-\hat{a})}|N\rangle ,  \notag \\
|g_{2},N^{(+)}\rangle &\equiv &|g_{2}\rangle \otimes e^{-(\mu /\omega )(\hat{%
a}^{\dag }-\hat{a})}|N\rangle ,  \label{eg1}
\end{eqnarray}%
with energies%
\begin{eqnarray}
E_{eN} &=&\varepsilon _{e}+\hbar \omega (N+\frac{1}{2}),  \notag \\
E_{g_{1}N} &=&E_{g_{2}N}=\varepsilon _{g}+\hbar \omega (N+\frac{1}{2})-\hbar 
\frac{\mu ^{2}}{\omega }.  \label{egen1}
\end{eqnarray}%
Hear $D\left( \alpha \right) =e^{\alpha (\hat{a}^{\dag }-\hat{a})}$ is the
displacement operator and quantum number $N=0,1,...$. The states $%
|N^{(+)}\rangle $, $|N^{(-)}\rangle $ are position-displaced Fock states:%
\begin{eqnarray}
|N^{(+)}\rangle &=&e^{-(\mu /\omega )(\hat{a}^{\dag }-\hat{a})}|N\rangle
=\sum_{M}I_{N,M}\left( \frac{\mu ^{2}}{\omega ^{2}}\right) |M\rangle , 
\notag \\
|N^{(-)}\rangle &=&e^{(\mu /\omega )(\hat{a}^{\dag }-\hat{a})}|N\rangle
=\sum_{M}I_{M,N}\left( \frac{\mu ^{2}}{\omega ^{2}}\right) |M\rangle ,
\label{PDFock}
\end{eqnarray}%
where $I_{N,M}\left( \alpha \right) $ is the Lagger function and defined via
generalized Lagger polynomials $L_{n}^{l}\left( \alpha \right) $ as follows:%
\begin{eqnarray}
I_{s,s^{\prime }}\left( \alpha \right) &=&\sqrt{\frac{s^{\prime }!}{s!}}e^{-%
\frac{\alpha }{2}}\alpha ^{\frac{s-s^{\prime }}{2}}L_{s^{\prime
}}^{s-s^{\prime }}\left( \alpha \right) =\left( -1\right) ^{s-s^{\prime
}}I_{s^{\prime },s}\left( \alpha \right) ,  \notag \\
L_{n}^{l}\left( \alpha \right) &=&\frac{1}{n!}e^{\alpha }\alpha ^{-l}\frac{%
d^{n}}{d\alpha ^{n}}\left( e^{-\alpha }\alpha ^{n+l}\right) .  \label{Lag}
\end{eqnarray}%
Particularly, $|0^{(+)}\rangle $ and $|0^{(-)}\rangle $ are the Glauber or
coherent states with mean number of photons $\mu ^{2}/\omega ^{2}$. Thus, we
have three ladders, two of them are crossed, and one ladder shifted by the
energy:%
\begin{equation}
\hbar \omega _{eg}=\hbar \left( \omega _{0}+\mu ^{2}/\omega \right) ,
\label{shift}
\end{equation}%
where $\omega _{0}=\left( \varepsilon _{e}-\varepsilon _{g}\right) /\hbar $.
The coupling term (\ref{h2}) $\widehat{V}\sim \widehat{S}_{t}$ induces
transitions between these manifolds. At the resonance:%
\begin{equation}
\omega _{eg}-\omega n=\delta _{n};\ \left\vert \delta _{n}\right\vert
<<\omega  \label{res}
\end{equation}%
with $n=1,2,...$ the equidistant ladders are crossed: $E_{eN}\simeq
E_{g_{1}N+n}=E_{g_{2}N+n}$, and the energy levels starting from the ground
state of upper harmonic oscillators are nearly threefold degenerated. The
coupling (\ref{h2}) removes this degeneracy, leading to "qutrit-photon"
entangled states. The splitting of levels is defined by the vacuum
multiphoton Rabi frequency. In this case we should apply secular
perturbation theory \cite{Landau}. Taking into account that 
\begin{equation*}
\langle g_{1},N^{(-)}|\widehat{V}|e,N-n\rangle =\left( -1\right) ^{n}\langle
g_{2},N^{(+)}|\widehat{V}|e,N-n\rangle ,
\end{equation*}%
and searching for the solution in the form%
\begin{equation*}
|\alpha ,N\rangle =C_{g_{1}}^{(\alpha )}|g_{1},N^{(-)}\rangle
+C_{g_{2}}^{(\alpha )}|g_{2},N^{(+)}\rangle
\end{equation*}%
\begin{equation}
+C_{e}^{(\alpha )}|e,N-n\rangle ,  \label{psi}
\end{equation}%
we get eigenenergies 
\begin{equation}
E_{1,N}=\varepsilon _{g}+\hbar \omega (N+\frac{1}{2})-\hbar \frac{\mu ^{2}}{%
\omega },  \label{E1}
\end{equation}%
\begin{equation}
E_{2,N}=E_{1,N}+\sqrt{2}\left\vert V_{N}\left( n\right) \right\vert ,
\label{E2}
\end{equation}%
\begin{equation}
E_{3,N}=E_{1,N}-\sqrt{2}\left\vert V_{N}\left( n\right) \right\vert ,
\label{E3}
\end{equation}%
and corresponding eigenstates 
\begin{equation}
|1,N\rangle =\frac{1}{\sqrt{2}}\left( |g_{1},N^{(-)}\rangle +\left(
-1\right) ^{n+1}|g_{2},N^{(+)}\rangle \right) ,  \label{psi1}
\end{equation}%
\begin{eqnarray}
|2,N\rangle &=&\frac{1}{2}|g_{1},N^{(-)}\rangle +\left( -1\right) ^{n}\frac{1%
}{2}|g_{2},N^{(+)}\rangle  \notag \\
&&+\frac{e^{-i\varphi _{V_{N}}}}{\sqrt{2}}|e,N-n\rangle ,  \label{psi2}
\end{eqnarray}%
\begin{eqnarray}
|3,N\rangle &=&\frac{1}{2}|g_{1},N^{(-)}\rangle +\left( -1\right) ^{n}\frac{1%
}{2}|g_{2},N^{(+)}\rangle  \notag \\
&&-\frac{e^{-i\varphi _{V_{N}}}}{\sqrt{2}}|e,N-n\rangle .  \label{psi3}
\end{eqnarray}%
In Eqs. (\ref{E2})-(\ref{psi3}) the transition matrix element is:%
\begin{equation*}
V_{N}\left( n\right) \equiv \langle g_{1},N^{(-)}|\widehat{V}|e,N-n\rangle
\end{equation*}%
\begin{equation*}
=\hbar \lambda \sqrt{N-n}I_{N-n-1,N}\left( \frac{\mu ^{2}}{\omega ^{2}}%
\right)
\end{equation*}%
\begin{equation}
+\hbar \lambda \sqrt{N-n+1}I_{N-n+1,N}\left( \frac{\mu ^{2}}{\omega ^{2}}%
\right) ,  \label{Vn}
\end{equation}%
and $\varphi _{V_{N}}=\arg \left[ V_{N}\left( n\right) \right] $. Thus,
starting from the level $N=n$ we have qutrit-photon entangled states (\ref%
{psi2}) and (\ref{psi3}), while for $N=0,1...n-1$ we have twofold
degenerated eigenenergies $E_{1N}$ with states $|g_{1},N^{(-)}\rangle $ and $%
|g_{2},N^{(+)}\rangle $. At the $\mu =0$ similar to the conventional JC
model there is a selection rule: $V_{N}\left( n\right) \neq 0$ only for $%
n=\pm 1$. In this case only one photon Rabi oscillations takes place. In our
model with $\mu \neq 0$ there are transition with arbitrary $n$ giving rise
to multiphoton coherent transitions. The solutions (\ref{psi1})-(\ref{psi3})
are valid at near multiphoton resonance $\omega _{eg}\simeq n\omega $ and
weak coupling: 
\begin{equation}
\left\vert V_{N}\left( n\right) \right\vert <<\hbar \omega .  \label{app}
\end{equation}

\section{Multiphoton Rabi Oscillations in the Ultrastrong Coupling Regime}

In this section, we consider temporal evolution of the qutrit-photonic field
system. This is of particular interest for applications in quantum
information processing. Here we also present numerical solutions of the
time-dependent Schr\"{o}dinger equation with the full Hamiltonian (\ref{H_m}%
).

We first proceed to consider the quantum dynamics of the coupled
qutrit-photonic field starting from an initial state, which is not an
eigenstate of the Hamiltonian (\ref{H_m}). Assuming arbitrary initial state $%
|\Psi _{0}\rangle $ of a system, then the state vector for times $t>0$ is
just given by the expansion over the basis obtained above:%
\begin{equation*}
|\Psi \left( t\right) \rangle =\sum_{N=0}^{n-1}\langle g_{1},N^{(-)}||\Psi
_{0}\rangle e^{-\frac{i}{\hbar }E_{g_{1}N}t}|g_{1},N^{(-)}\rangle
\end{equation*}%
\begin{equation*}
+\sum_{N=0}^{n-1}\langle g_{2},N^{(+)}||\Psi _{0}\rangle e^{-\frac{i}{\hbar }%
E_{g_{2}N}t}|g_{2},N^{(+)}\rangle
\end{equation*}%
\begin{equation}
+\sum_{\alpha =1}^{3}\sum_{N=n}^{\infty }\langle \alpha ,N||\Psi _{0}\rangle
e^{-\frac{i}{\hbar }E_{\alpha ,N}t}|\alpha ,N\rangle .  \label{DWF}
\end{equation}

For concreteness we will consider two common initial conditions for photonic
field: the Fock state and the coherent state. We will calculate the time
dependence of the three-level system population inversion%
\begin{equation}
W_{n}\left( t\right) =\langle \Psi \left( t\right) |\widehat{\Sigma }%
_{z}|\Psi \left( t\right) \rangle  \label{Wn}
\end{equation}%
at the exact $n$-photon resonance (\ref{res}), where 
\begin{equation}
\widehat{\Sigma }_{z}=\left( 
\begin{array}{ccc}
-1 & 0 & 0 \\ 
0 & -1 & 0 \\ 
0 & 0 & 1%
\end{array}%
\right) .  \label{Sz}
\end{equation}%
For the field in the vacuum state and two level system in the excited state $%
|\Psi _{0}\rangle =|e,0\rangle $, from Eqs. (\ref{psi1})-(\ref{psi3}), and (%
\ref{DWF}) we have:

\begin{equation}
|\Psi \left( t\right) \rangle =\frac{e^{i\varphi _{V_{n}}}}{\sqrt{2}}e^{-%
\frac{i}{\hbar }E_{2,n}t}\left( |2,n\rangle -e^{i\Omega _{n}\left( n\right)
t}|3,n\rangle \right) ,  \label{psiR}
\end{equation}%
where 
\begin{equation}
\Omega _{N}\left( n\right) =\frac{2\sqrt{2}\left\vert V_{N}\left( n\right)
\right\vert }{\hbar }  \label{Rabi}
\end{equation}%
is the multiphoton vacuum Rabi frequency. From Eqs. (\ref{Wn})-(\ref{psiR})
for the population inversion we obtain%
\begin{equation}
W_{n}\left( t\right) =\cos \left( \Omega _{n}\left( n\right) t\right) ,
\label{inv}
\end{equation}%
which corresponds to Rabi oscillations with periodic exchange of $n$ photons
between the qutrit and the radiation field.

Then we turn to the case in which a qutrit begins in the excited state, with
a photonic field prepared in a coherent state with a mean photon number $%
\overline{N}$:%
\begin{equation}
|\Psi _{0}\rangle =|e\rangle \otimes e^{\sqrt{\overline{N}}(\hat{a}^{\dag }-%
\hat{a})}|0\rangle .  \label{in0}
\end{equation}%
Taking into account Eqs. (\ref{DWF}) and (\ref{in0}) for the wave function
we obtain 
\begin{equation*}
\ |\Psi \left( t\right) \rangle =\sum_{N=n}^{\infty }\frac{e^{i\varphi
_{V_{N}}}}{\sqrt{2}}I_{N-n,0}\left( \overline{N}\right) \left[ e^{-\frac{i}{%
\hbar }E_{2,N}t}|2,N\rangle \right.
\end{equation*}%
\begin{equation}
\left. -e^{-\frac{i}{\hbar }E_{3,N}t}|3,N\rangle \right] ,  \label{Psit}
\end{equation}%
which in turn for population inversion (\ref{Wn}) gives:

\begin{equation}
W_{n}\left( t\right) =\sum_{N=0}^{\infty }\frac{e^{-\overline{N}}}{N!}%
\overline{N}^{N}\cos \left[ \Omega _{N+n}\left( n\right) t\right] .
\label{Wnt}
\end{equation}%
In this case we have superposition of Rabi oscillations with the amplitudes
given by the Poissonian distribution $P_{N}=e^{-\overline{N}}\overline{N}%
^{N}/N!$. As a consequence, we have collapse and revival phenomena of the
multiphoton Rabi oscillations. There are dominant frequencies in Eq. (\ref%
{Wnt}) as a result of the spread of probabilities about $\overline{N}$ for a
photon numbers in the range $\overline{N}\pm \sqrt{\overline{N}}$. When
these terms are oscillating out of phase with each other in the sum (\ref%
{Wnt}), it is expected cancellation of these terms, i.e. collapse of Rabi
oscillations. Hence, for large photon numbers $\overline{N}>>\sqrt{\overline{%
N}}$ the collapse time may be estimated as%
\begin{equation}
t_{\mathrm{c}}^{(n)}\simeq \frac{\pi }{2\sqrt{\overline{N}}}\left( \frac{%
\partial \Omega _{N}\left( n\right) }{\partial N}\right) ^{-1}.  \label{colt}
\end{equation}%
Taking into account Eq. (\ref{Vn}), it follows that in contrast to
conventional JC model the collapse time (\ref{colt}) strongly depends on the
mean photon number. 
\begin{figure}[tbp]
\includegraphics[width=.48\textwidth]{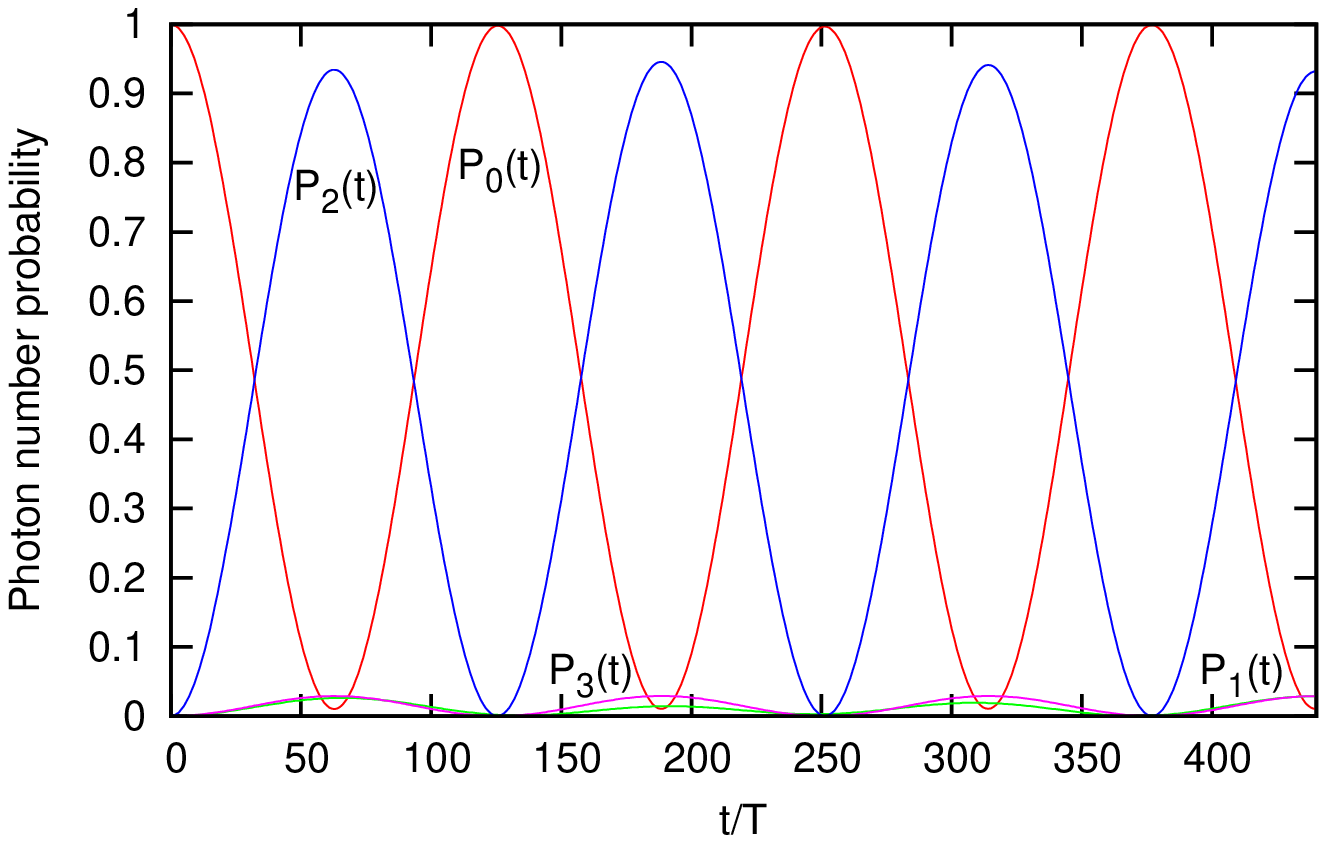}
\caption{(Color online) Photon number probability $P_{N}\left( t\right) $ as
a function of scaled time at the two-photon resonance ($2\protect\omega =%
\protect\omega _{eg}$). $\protect\lambda /\protect\omega =0.02$,$\ \protect%
\mu /\protect\omega =0.1$. }
\end{figure}
\begin{figure}[tbp]
\includegraphics[width=.47\textwidth]{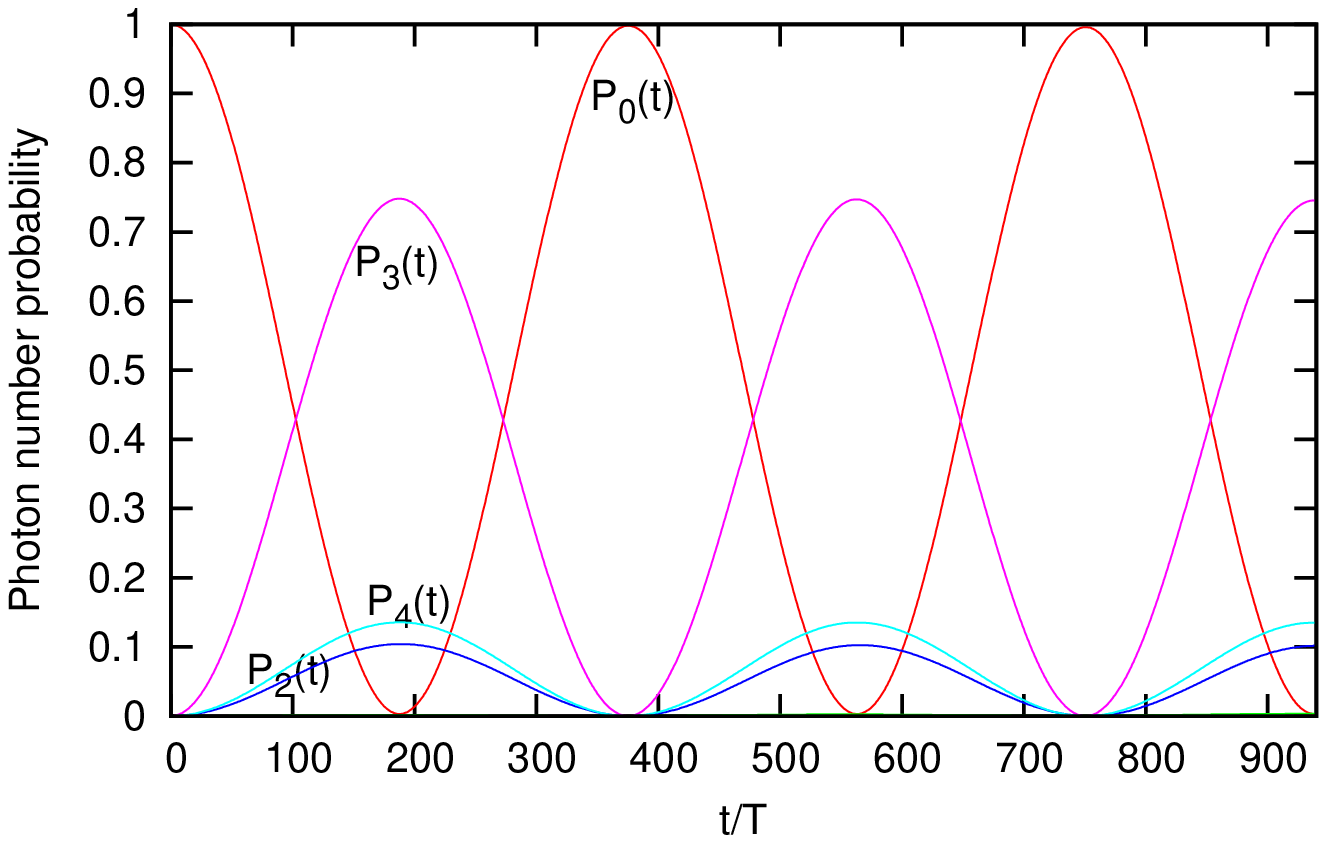}
\caption{(Color online) Photon number probability $P_{N}\left( t\right) $ as
a function of scaled time at the three-photon resonance ($n=3$). $\protect%
\lambda /\protect\omega =0.02$,$\ \protect\mu /\protect\omega =0.2$.}
\end{figure}
\begin{figure}[tbp]
\includegraphics[width=.47\textwidth]{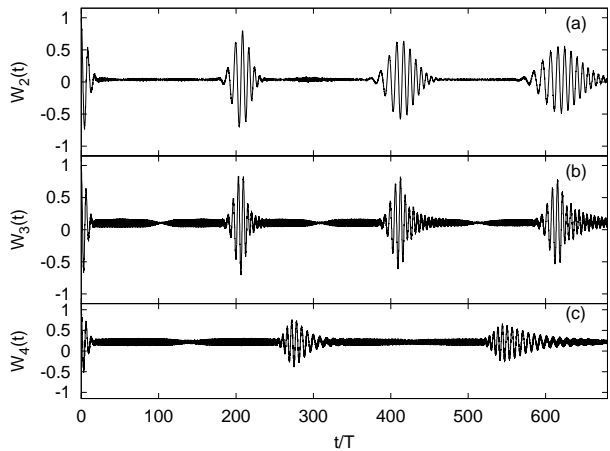}
\caption{Collapse and revival of the multiphoton Rabi oscillations.
Three-level system population inversion is shown with the field initially in
a coherent state. (a) Two-photon resonance with coupling parameters $\protect%
\lambda /\protect\omega =0.02$,$\ \protect\mu /\protect\omega =0.1$ and mean
photon number $\overline{N}=20$. (b) Three-photon resonance with parameters $%
\protect\lambda /\protect\omega =0.02$,$\ \protect\mu /\protect\omega =0.2$
and mean photon number $\overline{N}=30$. (c) Same as (b) but for
four-photon resonance and $\overline{N}=50$.}
\label{appwe}
\end{figure}
Now let us consider numerical solutions of the time dependent Schr\"{o}%
dinger equation with the full Hamiltonian (\ref{H_m}) in the Fock basis:%
\begin{equation}
|\Psi \left( t\right) \rangle =\sum_{\sigma
=g_{1},g_{2},e}\sum_{N=0}^{N_{\max }}C_{\sigma ,N}\left( t\right) |\sigma
\rangle \otimes |N\rangle .  \label{molt}
\end{equation}%
The set of equations for the probability amplitudes $C_{\sigma ,N}\left(
t\right) $ has been solved using a standard fourth-order Runge--Kutta
algorithm \cite{NR}, considering up to $N_{\max }=200$ excitations. To show
periodic multiphoton exchange between the qutrit and the radiation field, we
have calculated the population inversion (\ref{Wn}) and the photon number
probability:%
\begin{equation}
P_{N}\left( t\right) =\sum_{\sigma =g_{1},g_{2},e}\langle \sigma ,N||\Psi
\left( t\right) \rangle \langle \Psi \left( t\right) |\sigma ,N\rangle .
\label{Pn}
\end{equation}%
In Figs. (2) and (3) the photon number probability $P_{N}\left( t\right) $
as a function of time is shown for the two and three photon resonances. For
an initial state we assume qutrit in the excited state and the field in
vacuum state - $|e\rangle \otimes |0\rangle $. The Schr\"{o}dinger equation
with the full Hamiltonian (\ref{H_m}) was numerically solved with the
tunneling parameter $\Delta =0$. As is seen from these figures, due to the
mean dipole moment, multiphoton Fock states are excited. Figure 4 displays
collapse and revival of the multiphoton Rabi oscillations. Here the qutrit
population inversion is shown with the field initially in a coherent state
at two-, three-, and four-photon resonances for different mean photon
numbers.

The consequence of collapse and revival of the multiphoton Rabi oscillations
on the statistical properties of the photons shown in Fig. (5). For this
propose we have calculated the Mandel's $Q$-factor defined as \cite{Mand}:%
\begin{equation}
Q=\frac{\overline{N^{2}}-\overline{N}^{2}-\overline{N}}{\overline{N}}.
\label{Q}
\end{equation}%
When $-1\leq Q<0$ ($Q>0$), the statistics is sub-Poissonian
(super-Poissonian) and $Q=0$ shows the Poissonian statistics, which takes
place for coherent state. As is seen from this figure during the collapse
and revival of the multiphoton Rabi oscillations photons' antibunching ($Q<0$%
) takes place.

Concluding, we see that the numerical simulations are in agreement with
analytical treatment in the multiphoton resonant approximation and confirm
the revealed physical picture described above. 
\begin{figure}[tbp]
\includegraphics[width=.47\textwidth]{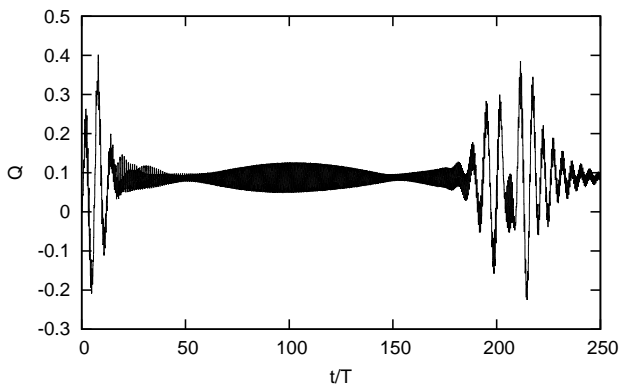}
\caption{Mandel's $Q$-factor versus scaled time for the setup of Fig. 4(b).}
\end{figure}

\section{Deep Strong Coupling Regime}

In this section we present results of numerical calculations that
demonstrate the properties of the system in the deep strong coupling regime.
In particular, we perform calculations for the ground state of the system.

The results obtained in the previous section rely on a resonant
approximation, which is valid for not too strong coupling (\ref{app}).
However, for certain experimental conditions, interaction strengths can
reach deep strong coupling regime. For those couplings, the application of a
resonant approximation is not justified anymore. Hence, one needs new
methods to examine the system in this parameter range. The Hamiltonian may
be written in matrix form in the basis $|g_{1},N\rangle $, $|g_{2},N\rangle $%
, and $|e,N\rangle $ (where $N=0,1,2,\dots $), which is the eigenbasis of
the noninteracting Hamiltonian with $\Delta =0$: 
\begin{widetext}
\begin{equation}
\widehat{H}=\left[ 
\begin{array}{cccccccccc}
\varepsilon _{g}+\frac{\hbar \omega }{2} & \Delta  & 0 & -\hbar \mu  & 0 & 
\hbar \lambda  & 0 & 0 & 0 & \dots  \\ 
\Delta  & \varepsilon _{g}+\frac{\hbar \omega }{2} & 0 & 0 & \hbar \mu  & 
-\hbar \lambda  & 0 & 0 & 0 & \dots  \\ 
0 & 0 & \varepsilon _{e}+\frac{\hbar \omega }{2} & \hbar \lambda  & -\hbar
\lambda  & 0 & 0 & 0 & 0 & \dots  \\ 
-\hbar \mu  & 0 & \hbar \lambda  & \varepsilon _{g}+\frac{3\hbar \omega }{2}
& \Delta  & 0 & -\sqrt{2}\hbar \mu  & 0 & \sqrt{2}\hbar \lambda  & \dots  \\ 
0 & \hbar \mu  & -\hbar \lambda  & \Delta  & \varepsilon _{g}+\frac{3\hbar
\omega }{2} & 0 & 0 & \sqrt{2}\hbar \mu  & -\sqrt{2}\hbar \lambda  & \dots 
\\ 
\hbar \lambda  & -\hbar \lambda  & 0 & 0 & 0 & \varepsilon _{e}+\frac{3\hbar
\omega }{2} & \sqrt{2}\hbar \lambda  & -\sqrt{2}\hbar \lambda  & 0 & \dots 
\\ 
0 & 0 & 0 & -\sqrt{2}\hbar \mu  & 0 & \sqrt{2}\hbar \lambda  & \varepsilon
_{g}+\frac{5\hbar \omega }{2} & \Delta  & 0 & \dots  \\ 
0 & 0 & 0 & 0 & \sqrt{2}\hbar \mu  & -\sqrt{2}\hbar \lambda  & \Delta  & 
\varepsilon _{g}+\frac{5\hbar \omega }{2} & 0 & \dots  \\ 
0 & 0 & 0 & \sqrt{2}\hbar \lambda  & -\sqrt{2}\hbar \lambda  & 0 & 0 & 0 & 
\varepsilon _{e}+\frac{5\hbar \omega }{2} & \dots  \\ 
\vdots  & \vdots  & \vdots  & \vdots  & \vdots  & \vdots  & \vdots  & \vdots 
& \vdots  & \ddots 
\end{array}%
\right] ,  \label{WH}
\end{equation}%
\end{widetext}where the order of the columns and rows is $|g_{1},0\rangle $, 
$|g_{2},0\rangle $, $|e,0\rangle $, $|g_{1},1\rangle $, $|g_{2},1\rangle $, $%
|e,1\rangle ...$. The Hamiltonian represents a block tridiagonal matrix for
which there are effective algorithms for diagonalization. Using the Arnoldi
algorithm \cite{WM} we have diagonalized the Hamiltonian (\ref{WH}). For
these calculations the energy is counted from $\varepsilon _{g}+\hbar \omega
/2$. In Figs. 6 and 7 we plot energy levels as a function of the scaled
coupling strength $\lambda /\omega $ at the two- and three-photon resonances
for the moderately strong coupling $\mu /\omega =0.3$. As is seen starting
from the level $n=\omega _{0}/\omega $ threefold degenerated states are
splitted and splitting energy increases for the large coupling strength in
accordance with resonant approximation.

\begin{figure}[tbp]
\includegraphics[width=.48\textwidth]{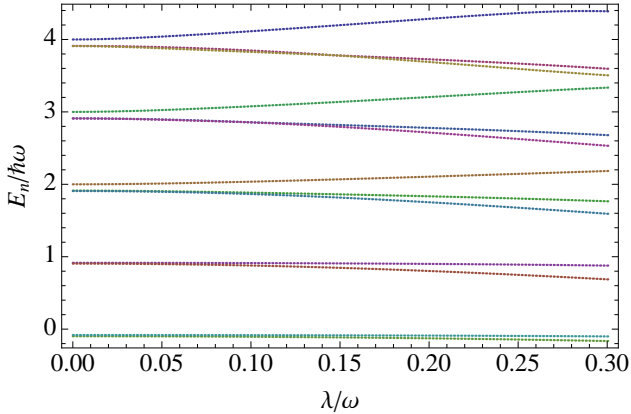}
\caption{(Color online) Lowest thirteen energy levels at the two-photon
resonance: $\protect\omega _{0}=2\protect\omega $. The scaled energy $%
E_{n}/\hbar \protect\omega $ is plotted as a function of the scaled coupling
strength $\protect\lambda /\protect\omega $ at $\protect\mu /\protect\omega %
=0.3$ and $\Delta /\hbar \protect\omega =0.01$. The energy is counted from $%
\protect\varepsilon _{g}+\hbar \protect\omega /2$.}
\end{figure}

\begin{figure}[tbp]
\includegraphics[width=.48\textwidth]{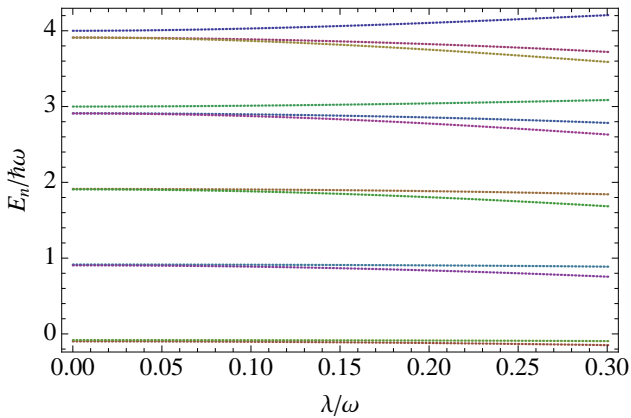}
\caption{(Color online) Lowest twelve energy levels at the three-photon
resonance: $\protect\omega _{0}=3\protect\omega $. The scaled energy $%
E_{n}/\hbar \protect\omega $ is plotted as a function of the scaled coupling
strength $\protect\lambda /\protect\omega $ at $\protect\mu /\protect\omega %
=0.3$ and $\Delta /\hbar \protect\omega =0.01$. }
\end{figure}

\begin{figure}[tbp]
\includegraphics[width=.48\textwidth]{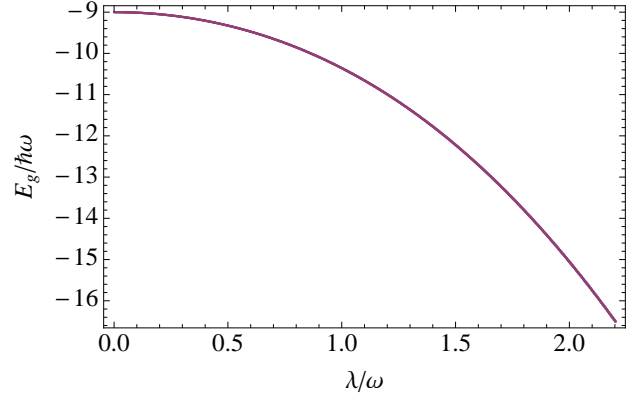}
\caption{The ground state scaled energy versus scaled coupling strength $%
\protect\lambda /\protect\omega $ at $\protect\mu /\protect\omega =3$ and $%
\protect\omega _{0}/\protect\omega =10$. }
\end{figure}

While in the JC model the ground state of the atom-photonic field system
consists of a product of the atom's ground state and the photonic field's
vacuum state, an inclusion of the terms $\sim \mu $ leads to a vacuum state (%
\ref{eg1})\ containing photons in the coherent states. With the large
coupling $\lambda $ one can expect\ qutrit-photonic field entangled ground
state containing large number of photons in the coherent states. The number
of photons will depend on both coupling parameters $\mu $ and $\lambda $. In
Fig. 8 it is plotted the ground state energy versus scaled coupling strength 
$\lambda /\omega $ for large $\mu $ and $\omega _{0}$ (henceforth we set $%
\Delta /\hbar \omega =0.1$). In Figs. 9 and 10 we show photon number
probability distribution $P_{N}$ in the ground state at $\lambda /\omega $ $%
=1$ and $\lambda /\omega $ $=2$, respectively. As is seen from this figures
the mean number of photons in the ground state is quite large and strongly
depends on the coupling $\lambda $ between qutrit excited and ground states.
For both setups Mandel's $Q$-factor (\ref{Q}) is calculated to be $Q\simeq
10^{-2}$, which means that for large $\mu $ and $\lambda $ photons exhibit
the Poissonian statistics.

\begin{figure}[tbp]
\includegraphics[width=.48\textwidth]{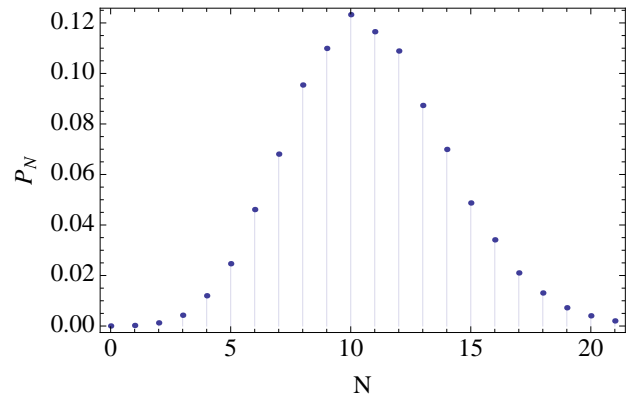}
\caption{Photon number probability distribution $P_{N}$ as a function of
photon number in the ground state at $\protect\omega _{0}/\protect\omega =10$%
, $\protect\mu /\protect\omega =3$, and $\protect\lambda /\protect\omega $ $%
=1$.}
\end{figure}

\begin{figure}[tbp]
\includegraphics[width=.48\textwidth]{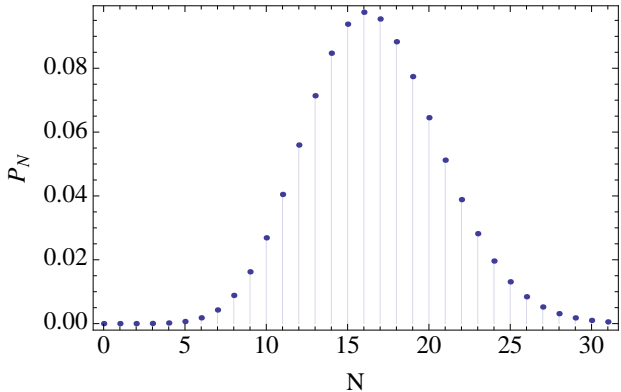}
\caption{Same as Fig. 9 but for $\protect\lambda /\protect\omega $ $=2$.}
\end{figure}

In order to obtain optimal conditions for the qutrit-oscillator entangled
states, we now analyze the entanglement properties in the ground state.
Various measures of entanglement exist. One commonly used entanglement
measures for pure states is the von Neumann entropy. The latter for the
qutrit will be defined as:%
\begin{equation}
S_{3}=-\mathrm{Tr}\left[ \rho _{r}\log _{3}\rho _{r}\right] ,  \label{S3}
\end{equation}%
where%
\begin{equation}
\rho _{r}=\mathrm{Tr}_{photon}\left[ |\Psi \rangle \langle \Psi |\right]
\label{reduce}
\end{equation}%
is the qutrit reduced density matrix and $|\Psi \rangle $ is the wave
function of qutrit-photon field combined system. The von Neumann entropy
satisfies the inequality $0\leq S_{3}\leq 1$, where the lower bound is
reached if and only if $|\Psi \rangle $ is a product state, while upper
bound is reached if and only if $|\Psi \rangle $ is a maximally entangled
state. Thus, rising from the Hamiltonian (\ref{WH}) we calculate the ground
state eigenvector of the combined system and then evaluate the entropy of
that state according to Eqs. (\ref{reduce}) and (\ref{S3}). In Figs. 11 and
12 the qutrit's entropy $S_{3}$, which quantifies the qutrit-photon field
entanglement in the ground state, is displayed as a function of coupling
parameters. Figure 11 is plotted for the fixed $\mu $, while Fig. 12 for the
fixed $\lambda $. As is seen from last two figures, there are optimal values
for the maximal entanglment and for very large couplings the latter vanishes.

\begin{figure}[tbp]
\includegraphics[width=.48\textwidth]{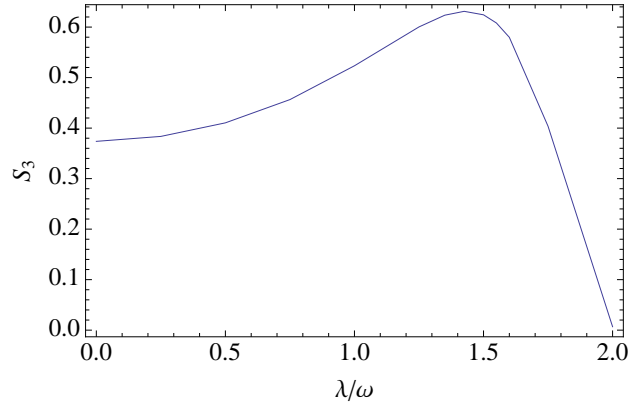}
\caption{The qutrit's entropy $S_{3}$ in the ground state as a function of $%
\protect\lambda /\protect\omega $ at $\protect\omega _{0}/\protect\omega =10$
and $\protect\mu /\protect\omega =3$.}
\end{figure}

\begin{figure}[tbp]
\includegraphics[width=.48\textwidth]{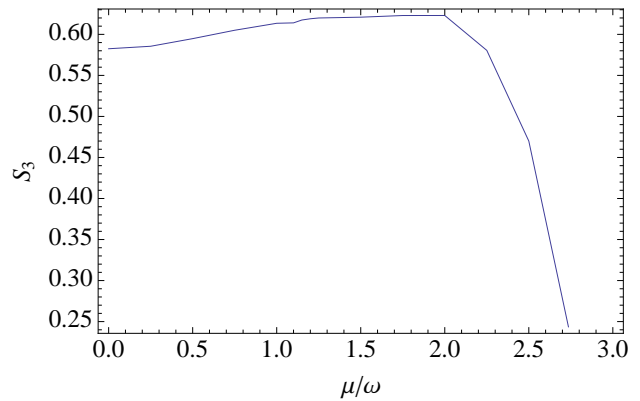}
\caption{The qutrit's entropy $S_{3}$ in the ground state as a function of $%
\protect\mu /\protect\omega $ at $\protect\omega _{0}/\protect\omega =10$
and $\protect\lambda /\protect\omega =2$.}
\end{figure}

\section{Conclusion}

We have presented a theoretical treatment of the quantum dynamics of a
qutrit in a polar-$\Lambda $ configuration interacting with a single-mode
photonic field in the ultrastrong and deep strong coupling regimes. For the
ultrastrong couplings we have solved the Schr\"{o}dinger equation in the
multiphoton resonant approximation and obtained simple analytical
expressions for the eigenstates and eigenenergies. In this case for the $n$%
-photon resonance we have entangled states of a qutrit and
position-displaced Fock states. We have also investigated the temporal
quantum dynamics of the considered system at the multiphoton resonance and
showed that due to the mean dipole moments in the lower states it is
possible Rabi oscillations of population inversion with periodic multiphoton
exchange between a qutrit and a photonic field. For the quantized field
prepared initially in a coherent state multiphoton Rabi oscillations
collapse/revive and photons' antibunching takes place. In the deep strong
coupling regime particular emphasis is placed on the ground state of the
system. The latter exhibits strictly nonclassical properties. In particular,
it has been shown that for the large coupling parameters $\mu $ and $\lambda 
$ we have qutrit-photonic field entangled ground state containing large
number of photons in the coherent states. The proposed model may have
diverse applications in QED with artificial atoms, especially in the circuit
QED, where the considered qutrit configuration and \ deep strong coupling
regime are foreseen.

\begin{acknowledgments}
This work was supported by State Committee of Science of Republic of
Armenia, Project No. 13RF-002 and Russian Foundation for Basic Research,
Project No. 13-02-90600.
\end{acknowledgments}

\appendix

\section{Equivalence of polar-$\Lambda $ and $L$ configurations}

\label{sec:L}

In this Appendix we prove equivalence of polar-$\Lambda $\ and $L$\
configurations. As an illustrative physical system we consider an electron
in a symmetric 1D double well potential. In this case the selection rule for
optical transitions is: the matrix element of the electric dipole moment is
nonzero for the states of different parity. Consider the case when the first
two eigenstates are localized in the wells of the potential, while the
higher eigenstate is delocalized. This situation can also be realized for
flux qubits \cite{77}. Thus, the ground eigenstate is an even function, the
eigenstate corresponding to adjacent level is an odd function, and finally,
the eigenstate of the excited state is an even function. According to
selection rule we have $L$\ configuration and in the single mode photonic
field one can write the Hamiltonian:\textrm{\ }%
\begin{equation*}
\widehat{H}_{\mathrm{L+ph}}=\hbar \omega \left( \widehat{a}^{+}\widehat{a}+%
\frac{1}{2}\right) +\widehat{H}_{\mathrm{L}}
\end{equation*}%
\begin{equation}
+\epsilon d_{g_{1}g_{2}}\widehat{S}_{g_{1}\leftrightarrow g_{2}}\left( 
\widehat{a}^{+}+\widehat{a}\right) +\epsilon d_{g_{2}e}\widehat{S}%
_{g_{2}\leftrightarrow e}\left( \widehat{a}^{+}+\widehat{a}\right) ,
\label{A1}
\end{equation}%
where\textrm{\ }%
\begin{equation}
\widehat{H}_{\mathrm{L}}=\left( 
\begin{array}{ccc}
\varepsilon _{g_{1}} & 0 & 0 \\ 
0 & \varepsilon _{g_{2}} & 0 \\ 
0 & 0 & \varepsilon _{e}%
\end{array}%
\right) ,  \label{A2}
\end{equation}%
$d_{g_{1}g_{2}}$, $d_{g_{2}e}$\ are transition dipole moments, $\epsilon
\left( \widehat{a}^{+}+\widehat{a}\right) $\ is the electric field operator,
and\textrm{\ }%
\begin{equation}
\widehat{S}_{g_{1}\leftrightarrow g_{2}}=\left( 
\begin{array}{ccc}
0 & 1 & 0 \\ 
1 & 0 & 0 \\ 
0 & 0 & 0%
\end{array}%
\right) ,\widehat{S}_{g_{2}\leftrightarrow e}=\left( 
\begin{array}{ccc}
0 & 0 & 0 \\ 
0 & 0 & 1 \\ 
0 & 1 & 0%
\end{array}%
\right)  \label{A3}
\end{equation}%
are transition operators. Now let us apply unitary transformation ($\widehat{%
U}\widehat{L}\widehat{U}^{+}$) defined as:\textrm{\ }%
\begin{equation}
\widehat{U}=\frac{1}{\sqrt{2}}\left( 
\begin{array}{ccc}
1 & -1 & 0 \\ 
1 & 1 & 0 \\ 
0 & 0 & \sqrt{2}%
\end{array}%
\right) .  \label{A4}
\end{equation}%
For the transformed operators we obtain:\textrm{\ }%
\begin{equation}
\widehat{H}_{\mathrm{L}}^{\prime }=\left( 
\begin{array}{ccc}
\frac{\varepsilon _{g_{1}}+\varepsilon _{g_{2}}}{2} & \frac{\varepsilon
_{g_{1}}-\varepsilon _{g_{2}}}{2} & 0 \\ 
\frac{\varepsilon _{g_{1}}-\varepsilon _{g_{2}}}{2} & \frac{\varepsilon
_{g_{1}}+\varepsilon _{g_{2}}}{2} & 0 \\ 
0 & 0 & \varepsilon _{e}%
\end{array}%
\right) ,  \label{A5}
\end{equation}%
\begin{equation}
\widehat{S}_{g_{1}\leftrightarrow g_{2}}^{\prime }=\left( 
\begin{array}{ccc}
-1 & 0 & 0 \\ 
0 & 1 & 0 \\ 
0 & 0 & 0%
\end{array}%
\right) \equiv \widehat{S}_{L},  \label{A6}
\end{equation}%
\begin{equation}
\;\widehat{S}_{g_{2}\leftrightarrow e}^{\prime }=-\frac{1}{\sqrt{2}}\left( 
\begin{array}{ccc}
0 & 0 & 1 \\ 
0 & 0 & -1 \\ 
1 & -1 & 0%
\end{array}%
\right) \equiv -\frac{1}{\sqrt{2}}\widehat{S}_{t}.  \label{A7}
\end{equation}%
Now it is easy to see that the transformed Hamiltonian corresponds to polar-$%
\Lambda $\ configuration considered in the paper (see Eq. (\ref{H_m})) with
the parameters:\textrm{\ }%
\begin{eqnarray}
\varepsilon _{g} &=&\frac{\varepsilon _{g_{1}}+\varepsilon _{g_{2}}}{2}%
;\;\Delta =\frac{\varepsilon _{g_{1}}-\varepsilon _{g_{2}}}{2};  \notag \\
\mu &=&\frac{\epsilon d_{g_{1}g_{2}}}{\hbar };\;\lambda =-\frac{1}{\sqrt{2}}%
\frac{\epsilon d_{g_{2}e}}{\hbar }.  \label{A8}
\end{eqnarray}%
The terms $\widehat{S}_{L}$\ and $\widehat{S}_{t}$\ describe electric-dipole
moment matrix elements. So, the diagonal elements are the mean dipole
moments and are described by the terms proportional to $\widehat{S}_{L}$.
This is also obvious in the coordinate picture. In the symmetric double well
where the probability of tunnel transition between the "left" and "right"
potential wells is small ($\left\vert \Delta \right\vert <<\varepsilon _{g}$%
), we have nearly degenerated ground state and can use two equivalent bases.%
\textrm{\ }In a $L$\ configuration the two lowest energy eingenstates are of
the form $|\pm \rangle =\left( |\mathrm{left}\rangle \pm |\mathrm{right}%
\rangle \right) /\sqrt{2}$, where $|\mathrm{left}\rangle $\ and $|\mathrm{%
right}\rangle $\ are the basis wave functions in the polar-$\Lambda $\
configuration and represent the situations that the particle is in the left
or right potential well with opposite mean dipole moments.

\end{document}